\documentclass[pra,superscriptaddress,floatfix,twocolumn,showpacs]{revtex4}
\usepackage{amsmath,bbm,mathrsfs,amssymb,amssymb,graphicx,times,hyperref}
\usepackage[dvips]{color}
\newcommand{\tmmathbf}[1]{\ensuremath{\boldsymbol{#1}}}
\newcommand{\tmop}[1]{\ensuremath{\operatorname{#1}}}
\newcommand{\EO}{E_0}

\newcommand{\ED}{E_D}
\newcommand{\PO}{\Pi_0}
\newcommand{\PS}{\Pi_S}
\newcommand{\PD}{\Pi_D}
\newcommand{\RD}{R_D}
\newcommand{\one}{\mathbbm{I}}

\newcommand{\dket}[1]{|#1 {\rangle\!\rangle}}
\newcommand{\dbra}[1]{{\langle\!\langle} #1 |}
\newcommand{\dbrb}[1]{{\langle\!\langle} #1 }

\newcommand{\bmsigma}{\tmmathbf{{\sigma}}}
\newcommand{\bmsigmaOUT}{{\boldsymbol \sigma}_{\rm out}}
\newcommand{\psiOUT}{\boldsymbol\Psi_{\rm out}}
\newcommand{\bmS}{\tmmathbf{S}}
\newcommand{\bmLS}{\tmmathbf{LS}}
\newcommand{\bmSt}{\tmmathbf{S}}
\newcommand{\Sbs}{{\boldsymbol S}_{\rm BS}} 
\newcommand{\Ubs}{{U}_{\rm BS}}
\newcommand{\Id}{\mathbbm{1}}
\newcommand{\bra}[1]{\ensuremath{\left\langle #1
\right|}}\newcommand{\ket}[1]{\ensuremath{\left| #1
\right\rangle}}
\newcommand{\Tr}{\hbox{\rm Tr}}
\begin{document}
\title{Optimal unambiguous comparison of two unknown squeezed vacua}
\author{Stefano Olivares}
\email{stefano.olivares@mi.infn.it}
\affiliation{CNISM UdR Milano Universit\`a, I-20133 Milano, Italy}
\affiliation{Dipartimento di Fisica, Universit\`a degli Studi di Milano,
I-20133 Milano, Italy}
\author{Michal Sedl\'ak}
\email{fyzimsed@savba.sk}
\altaffiliation{On leave from Bratislava.}
\affiliation{Research Centre for Quantum Information, Institute of
  Physics, Slovak Academy of Sciences, D\'{u}bravsk\'{a} cesta 9, 845
  11 Bratislava, Slovak Republic}
\affiliation{QUIT group, Dipartimento di Fisica ``A. Volta'', via
  Bassi 6, I-27100 Pavia, Italy}
\author{Peter Rap\v{c}an}
\affiliation{Research Centre for Quantum
  Information, Institute of Physics, Slovak Academy of Sciences,
  D\'{u}bravsk\'{a} cesta 9, 845 11 Bratislava, Slovak Republic}
\author{Matteo G.~A.~Paris} \affiliation{Dipartimento di Fisica,
  Universit\`a degli Studi di Milano, I-20133 Milano, Italy}
\affiliation{CNISM UdR Milano Universit\`a, I-20133 Milano, Italy}
\author{Vladim\'\i r Bu\v{z}ek}
\affiliation{Research Centre for
  Quantum Information, Institute of Physics, Slovak Academy of
  Sciences, D\'{u}bravsk\'{a} cesta 9, 845 11 Bratislava, Slovak
  Republic} \affiliation{ Faculty of Informatics, Masaryk University,
  Botanick\'a 68a, 602 00 Brno, Czech Republic}
\begin{abstract}
  We propose a scheme for unambiguous state comparison (USC) of two
  unknown squeezed vacuum states of an electromagnetic field. Our
  setup is based on linear optical elements and photon-number
  detectors, and achieves optimal USC in an ideal case of unit
  quantum efficiency.  In realistic conditions, i.e., for non-unit
  quantum efficiency of photodetectors, we evaluate the probability of
  getting an ambiguous result as well as the reliability of the
  scheme, thus showing its robustness in comparison to previous
  proposals.
\end{abstract}
\maketitle
\section{Introduction}\label{s:intro}
The possibility of creating physical systems with identical properties
is crucial for any physical theory that is verifiable by experiments.
Comparison of preparators -- a procedure of determining whether they
prepare the same objects or not -- is one of the basic experiments we
would like to do when testing a theory, because it allows us to
operationally define equivalence of such devices for their further
use.  In the framework of classical physics, we can in principle
measure and determine the state of the system perfectly without
disturbing it. Thus, to compare states of two systems it suffices to
measure each system separately.  However, in quantum theory, due to
its statistical nature, we cannot make deterministic
conclusions/predictions even for the simplest experimental
situations. Therefore, the comparison of quantum states is different
compared to the classical situation.
\par
Imagine we are given two independently prepared quantum systems of the
same physical nature (e.g., two photons or two electrons). We would
like to determine unambiguously whether the (internal) states of these
two systems are the same or not. If we have just a single copy of each
of the states and we possess no further information about the
preparation then a measurement performed on each system separately
cannot determine the states precisely enough to allow an error-free
comparison. In this case, also all other strategies would fail,
because our knowledge about the states is insufficient
\cite{JexAnderson}, e.g., if each of the systems can be in an
arbitrary mixed state, then it is impossible to unambiguously test
whether the states are equal or not.  However, there are often
situations in which we have some additional \emph{a priori}
information on the states we want to compare. For example, we might
know that each system has been prepared in a pure state. This kind of
scenario has been considered in Ref.~\cite{comp1} for two qudits and
in Ref.~\cite{comp2} for the comparison of a larger number of
systems. Thereafter, the comparison of coherent states and its
application to quantum cryptography has been addressed in
Ref.~\cite{and:PRA:06}. Sedl\'ak {\em et al.}~\cite{rcqi1} analyzed
the comparison with more copies of the two systems and proposed an
optimal comparator for coherent states, which, on this subset,
outperforms the optimal universal comparator \cite{comp1} working for
all pure states.
\par
In the present paper we analyze the unambiguous quantum state
comparison (USC) of two unknown squeezed vacuum states, that is, we
would like to unambiguously determine whether two unknown
squeezed-vacuum states are the same or not. The conclusion has to be
drawn from a procedure using only a single copy of the states. At the
end of the procedure, using only the outcome of the measurement we
have to decide whether the two states given to us have been the same,
different, or that we don't know which of the former conclusions is
true. We strive to find an optimal procedure, i.e., one maximizing the
probability of correctly judging the equivalence of the compared
squeezed states.
\par
Our proposal relies on the interference of two squeezed states at a
beam splitter and on the subsequent measurement of the difference
between the number of detected photons at the two output ports. In
Ref.~\cite{and:PRA:06}, unambiguous comparison of coherent states has
been considered in detail and a short remark is devoted to the
comparison of squeezed vacua. In the setup of Ref.~\cite{and:PRA:06},
after interference at a beam splitter, one needs to measure the parity
of the detected number of photons: a detection of an odd number of
photons indicates the difference between the inputs. As a consequence,
the quantum efficiency of the detectors is a critical parameter and
plays a crucial role in the robustness of the scheme. As we will show,
this problem is less relevant in our case, since our setup requires
the measurement of the difference of the detected number of
photons. Our configuration also allows us to prove optimality of our
setup.
\par
The plan of the paper is as follows. In Section~\ref{s:setup} we
introduce our scheme to compare two squeezed vacuum states, whereas
the proof of the optimality of the setup is given in
Section~\ref{s:opt}. The performances of our scheme, also in the presence
of imperfections at the detection stage, are investigated in
Section~\ref{s:perform}, together with its reliability in the presence
of noise. Section~\ref{s:concl} closes the paper with concluding remarks.
\section{Comparison of squeezed vacuum states}\label{s:setup}
Our goal is the comparison of two squeezed vacuum states
$\ket{\xi}\equiv S(\xi)\ket{0}$ and $\ket{\zeta}\equiv
S(\zeta)\ket{0}$, where $S(\gamma) = \exp[\frac12\gamma (a^{\dag})^2 -
\frac12\gamma^* a^2]$ is the single-mode squeezing operator,
$\xi,\zeta,\gamma \in {\mathbbm C}$ \cite{Loudon}.  We let $\xi = r \,
e^{i\psi}$ and $\zeta = s\, e^{i\varphi}$, where $r= |\xi|$, $\psi =
\arg(\xi)$, $s=|\zeta|$, $\varphi = \arg(\zeta)$.
\begin{figure}[tb!]
  \includegraphics[width=0.98\columnwidth]{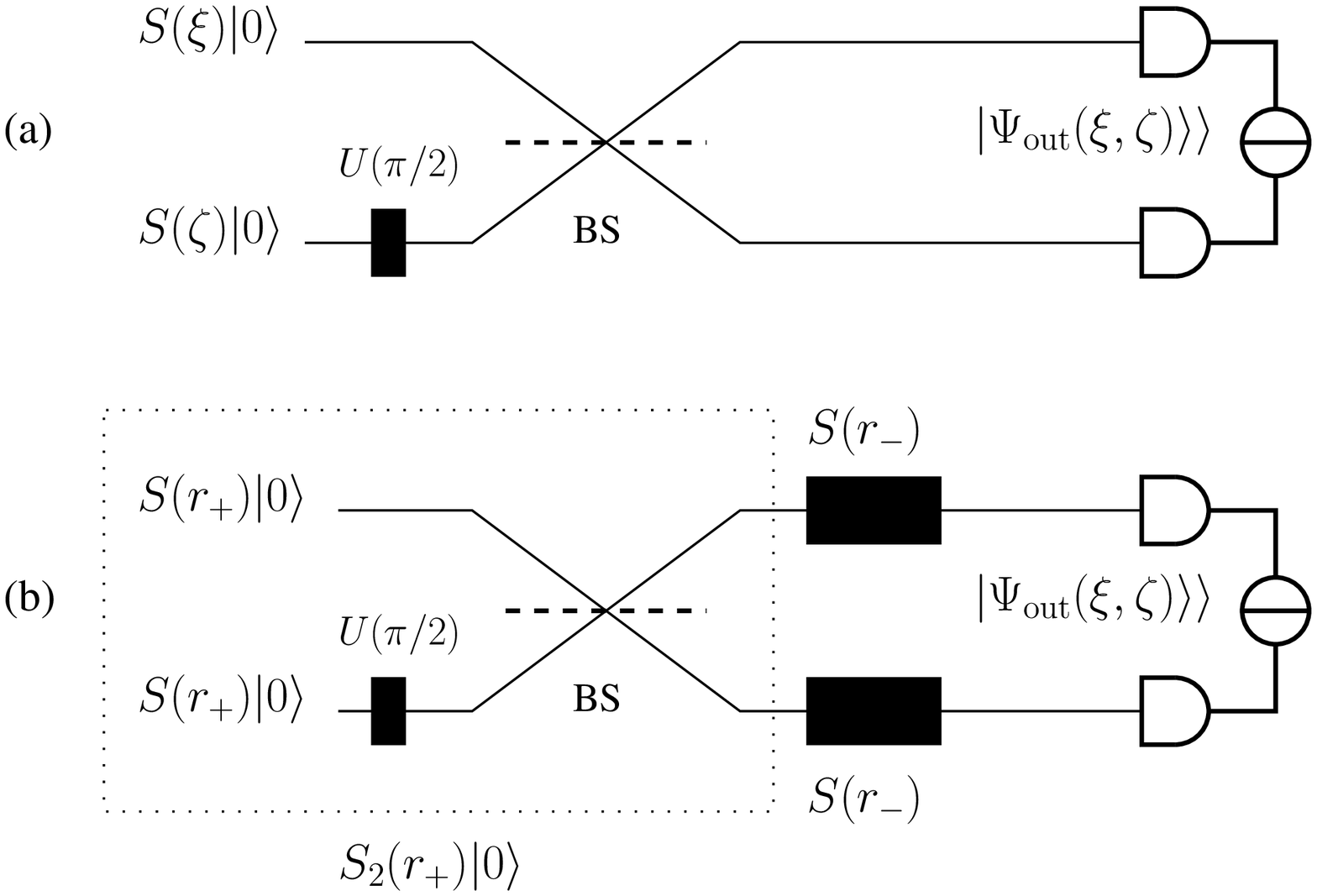}
  \caption{\label{f:schemes} (a) Schematic diagram of a setup for USC
    of the squeezed vacuum states $S (\xi) \ket{0}$ and $S (\zeta)
    \ket{0}$. (b) Scheme leading to the same output state
    $\dket{\Psi_{\rm out}(\xi,\zeta)}$ as in (a) when the
      two squeezing phases are unknown but equal, i.e.,
      $\arg(\xi)=\arg(\zeta)$. We defined $r_{\pm} = (\xi \pm\ \zeta)
    / 2$. See the text for comments and details.}
\end{figure}
We remind that a comparator is a measuring device with two systems at
the input and two or more possible outcomes, aimed at determining
whether the two systems have been prepared in the same state. The
setup we propose for the comparison of the two squeezed vacuum states
is composed of a phase shifter, beam splitter and photon-counting
detectors, and can be implemented with a current technology. The basic
idea is sketched in Fig.~\ref{f:schemes}~(a): we start from the two
squeezed vacuum states we wish to compare, $S (\xi) \ket{0}$ and $S
(\zeta) \ket{0}$.  At the first stage of our protocol, one of the two
states, say $S(\xi) \ket{0}$, undergoes a phase shift $U(\pi/2)$, i.e.
$U(\pi/2) S(\xi) \ket{0} = S(-\xi) \ket{0}$; then we mix the states,
having now orthogonal phases, at a balanced beam splitter (BS). If
$\xi=\zeta$, i.e., the input states are equal, then the output state
is the two-mode squeezed vacuum state of radiation (twin-beam state,
TWB) \cite{joint}, namely:
\begin{align} \label{TWB}
\dket{\psiOUT (\xi,\xi)}
&= U_{\rm BS}S(\xi)\otimes S(-\xi)\ket{0} \equiv S_2 (\xi) \ket{0}\\
&= \sqrt{1 - |\lambda(\xi)|^2}
\sum_{n = 0}^{\infty} \lambda(\xi)^n \ket{n} \ket{n},
\label{TWB:fock}
\end{align}
where $\ket{n} \ket{n}\equiv \ket{n} \otimes \ket{n}$, $U_{\rm BS}$ is
the unitary operator describing the action of the BS, $S_2 (\xi) =
\exp ( \xi a^{\dag} b^{\dag} - \xi^{*} ab )$ is the two-mode squeezing
operator acting on the two modes $a$ and $b$, respectively, and
$\lambda(\gamma) = e^{i\arg(\gamma)}\tanh |\gamma|$.  One finds
perfect correlations in the photon number of the two beams, that can
be detected, e.g., by measuring the difference between the number of
photons at the outputs (see Fig.~\ref{f:schemes}), that, in this case,
is always equal to zero. On the contrary, if $\xi\ne \zeta$, a
different number of photons can be detected in the two beams, as we
are going to show in the following.
\par
Though the setup in Fig.~\ref{f:schemes}~(a) works for generic $\xi$ and $\zeta$, in the following we address the scenario in which the two squeezing phases are unknown but equal, i.e., $\arg(\xi)=\arg(\zeta)$ [note that in Eq.~(\ref{TWB:fock}) a two-mode squeezed vacuum state results if and only if the two input squeezed states are the same, no matter the value of their phase]. This allows us to write the output state $\dket{\Psi_{\rm out}(\xi,\zeta)}$ in a simple form that will turn out to be useful to characterize our setup. Now, the same result of the evolution as in Fig.~\ref{f:schemes}~(a) can be obtained considering the scheme displayed in Fig.~\ref{f:schemes}~(b) (see Appendix \ref{app:CMevol}). Here the two input states with squeezing parameters $\xi$ and $\zeta$ are substituted with two squeezed vacuum states having the same squeezing parameter amplitude $r_{+} = (\xi + \zeta) / 2$. It is also worth noting that the scheme in Fig.~\ref{f:schemes}~(b) cannot be directly used for the comparison, since it requires a prior knowledge about the squeezing parameters to assess $r_{\pm}$. Now, after the mixing at the BS, the outgoing modes undergo two local squeezing operations with amplitude $r_{-} = (\xi - \zeta) / 2$. In formula one has the (formal) equivalence:
\begin{equation}
  \Ubs \hspace{0.25em} S (\xi) \otimes S (-\zeta) \ket{0} =
S (r_{-}) \otimes S (r_{-})
  \hspace{0.25em} S_2 (r_{+}) \ket{0} .
\end{equation}
Since
$S_2 (r_{+}) \ket{0} =
\sqrt{1 - |\lambda(r_{+})|^2} \sum_n \lambda(r_{+})^n \ket{n}\ket{n}$, we obtain:
\begin{equation}\label{eq:PsiOutReal}
  \dket{\psiOUT(\xi,\zeta)} = \sqrt{1 - |\lambda(r_{+})|^2} \sum_{n=0}^{\infty} \lambda(r_{+})^n
  \ket{\psi_n} \ket{\psi_n},
\end{equation}
where we defined the new basis $\ket{\psi_n} = S (r_{-})
\ket{n}$. Finally, the probability of measuring $h$ and $k$ photons in
the two beams, respectively, is given by:
\begin{equation}
p(h,k) = |\bra{h} \langle k \dket{\psiOUT(\xi,\zeta)}|^2,
\end{equation}
with:
\begin{align}\label{final:DPC}
  \bra{h} \langle k &\dket{\psiOUT(\xi,\zeta)} = \nonumber\\
  &\sqrt{1 - |\lambda(r_{+})|^2}\sum_n \lambda(r_{+})^n
  [S(r_{-})]_{hn} \hspace{0.25em} [S (r_{-})]_{kn},
\end{align}
where $[S (r_{-})]_{lm} = \bra{l} S (r_{-}) \ket{m}$ are the matrix
elements of the squeezing operator, whose analytical expressions are
given e.g., in Ref.~\cite{p:91}.  If $\xi=\zeta$ and $h\ne k$, then
$\bra{h} \langle k \dket{\psiOUT(r,r)} = 0$ and $p(h,k)=0$, as one can
see from Eq.~(\ref{TWB:fock}). Thus, the probability $p(h,k)$, for
$h\ne k$, can be non-zero only if $\xi\ne \zeta$, that is only if the
input states are different.
\par
In the ideal case (unit quantum efficiency of the detectors) the
measurement apparatus we want to use gives two possible outcomes: zero
or non-zero photon-counting difference. Thus, the POVM performed is
defined by the effects $\EO$ and $\ED$, corresponding to the ``zero''
and ``non-zero'' photon-counting events, respectively, given by:
\begin{align}
\label{POVM:comp} \EO= \sum^{\infty}_{n = 0}
\ket{n}\!\bra{n}\otimes \ket{n}\!\bra{n},\quad \ED =  \one - \EO.
\end{align}
The occurrence of the ``$D$'' event implies that the incident
squeezed-vacuum states could not have been identical [see
Eqs.~(\ref{eq:PsiOutReal}) and (\ref{final:DPC})].  The occurrence of
the ``$0$'' event, on the other hand, implies nothing, as each
possible pair of squeezed-vacuum states leads to a non-zero overlap
with any of the states $\ket{n} \ket{n}$.  Thus, event ``$D$''
unambiguously indicates the difference of the compared squeezed
states, whereas ``$O$'' is an inconclusive outcome.
\section{Proof of the optimality of the setup}\label{s:opt}
In this Section we prove optimality of the proposed setup for two situations: (i) the restricted scenario in which the squeezing phases of the compared states are unknown but equal for both states (ii) for the general situation, when no assumption on the squeezing phases is taken. We first tackle the former scenario considered in most of the paper and at the end of the proof we comment on the differences in proving (ii).

Let us denote by $\mathcal{S}^{\varphi}\equiv\{ S(r e^{i\varphi})
\ket{0}; r\in \mathbb{R} \}$ the set of squeezed states from which we
randomly chose the states to be compared. We also define the sets
$\mathcal{S}^{\varphi}_{S}\equiv\{ S(r e^{i\varphi}) \ket{0}\otimes
S(r e^{i\varphi})\ket{0}; r\in \mathbb{R} \}$,
$\mathcal{S}^{\varphi}_{D}\equiv
\mathcal{S}^{\varphi}\otimes\mathcal{S}^{\varphi}\backslash \,
\mathcal{S}^{\varphi}_{S}$, composed by pairs of identical and
different three outcomes ("same", "different" and "don't know")
described by the POVM $\PS+\PD+\PO=\mathbbm{I}$ and we optimize the
overall probability:
\begin{align}
P=&\,z_{S}\int_{\mathcal{S}^{\varphi}_{S}}
\mathrm{d}\Phi\; p_{S}(\Phi) \bra{\Phi}\PS\ket{\Phi} \nonumber\\
&+z_{D}\int_{\mathcal{S}^{\varphi}_{D}}
\mathrm{d}\Phi\; p_{D}(\Phi) \bra{\Phi}\PD\ket{\Phi},
\end{align}
where $z_D$ and $z_S=1-z_D$ are the a priori probability of being
different or the same, $p_{S}(\Phi)$, $p_{D}(\Phi)$ are probability
densities of choosing $\ket{\Phi}$ from $\mathcal{S}^{\varphi}_{S}$,
$\mathcal{S}^{\varphi}_{D}$, respectively.  We also impose the
no-error constraints:
\begin{subequations}\label{noerrorcond1}
\begin{align}
\Tr(\PS\ket{\Phi}\bra{\Phi})&=0, \quad
\forall \ket{\Phi}\in \mathcal{S}^{\varphi}_{D}, \\
\Tr(\PD\ket{\Phi}\bra{\Phi})&=0, \quad
\forall \ket{\Phi}\in \mathcal{S}^{\varphi}_{S},
\label{noerrorcond1:b}
\end{align}
\end{subequations}
which guarantee the unambiguity of the results.  From the mathematical point of view, the constraints (\ref{noerrorcond1}) restrict the support of the operators $\PS$ and $\PD$. The fact that the possible states in $\mathcal{S}^{\varphi}$ form a continuous subset of pure states, is responsible for the impossibility to unambiguously confirm that the compared states are identical. The proof of this statement can be found in Appendix \ref{app:nosame} and essentially states that, due to the no-error conditions (\ref{noerrorcond1}), we must have $\PS=0$. Thus, the measurement actually has only two outcomes, the effective POVM is given by $\PD, \PO=\one -\PD$, and it is clear that increasing the eigenvalues of $\PD$ without changing its support increases the figure of merit and leaves the no-error conditions satisfied. This is true independently of the distribution $p_{D}$ and thus the optimal measurement is formed by $\PD$ being a projector onto the biggest support allowed by the no-error condition (\ref{noerrorcond1}) and $\PO$ being a projector onto the orthocomplement. Moreover, the quantity that completely characterizes the behavior of the squeezed-states comparator is $p(D|r,s)=\bra{\Phi}\PD\ket{\Phi}$, i.e., the conditional probability of obtaining the outcome $\PD$ if different squeezed states $\ket{\Phi}=S(re^{i\varphi}) \ket{0}\otimes S(s e^{i\varphi})\ket{0}$ ($r\neq s$) are sent to the comparator. It is worth to note that in what follows one does not need to know the actual value of $\varphi$. Summarizing, in order to find an optimal comparator of squeezed states from $\mathcal{S}^{\varphi}$ we need to refine the definition of the largest allowed support of $\PD$ hidden in the no-error condition (\ref{noerrorcond1:b}). To do this we equivalently rewrite Eq.~(\ref{noerrorcond1:b}) as:
\begin{eqnarray}
\Tr(W \PD W^{\dagger} W\ket{\Phi}\bra{\Phi}W^{\dagger})&=&0 \quad
\forall \ket{\Phi}\in \mathcal{S}^{\varphi}_{S},
\end{eqnarray}
which, by denoting $\ED\equiv W \PD W^{\dagger}$ and choosing $W$ to
be the unitary transformation performed by the proposed setup from
Fig.~\ref{f:schemes}~(a), becomes:
\begin{eqnarray}
\Tr(\ED\dket{\Psi_{\rm out}(r,r)} \dbra{\Psi_{\rm out}(r,r)})&=&0
\quad
\forall r\in\mathbb{R}.
\label{noerrorcond2}
\end{eqnarray}
The optimality of the proposed setup is proved by showing that the
biggest support allowed by the previous condition coincides with the
support of the projective measurement $\ED$ we use, see
Eq.~(\ref{POVM:comp}).
\par
From the expression of $\dket{\Psi_{\rm out}(r,r)}$,
Eq.~(\ref{TWB:fock}), it is clear that for any operator $\ED$ with the
support orthogonal to the span of $\ket{n}\ket{n}$, with $n \in
{\mathbbm N}$, the unambiguous no-error condition (\ref{noerrorcond2})
holds. Hence, if any such operator $\ED$ is a part of a POVM, then the
emergence of the outcome related to it unambiguously indicates the
difference of the squeezing parameters. We now proceed to show that
the support of such $\ED$ cannot be further enlarged. Now let us
assume that a vector that a vector $\dket{v} = \sum_{h, k =
  0}^{\infty} d_{hk} \ket{h} \ket{k}$ with at least one non-zero
coefficient $d_{ii}$ is in the support of $\ED$. As a consequence of
the required no-error condition (\ref{noerrorcond2}) the overlap
\begin{equation}\label{overlap1}
\dbrb{v}\dket{\Psi_{\rm out} (r,r)} =
\sqrt{1 - |\lambda (r)|^2} \sum_{n = 0}^{\infty} d_{nn}^{\ast} \lambda (r)^n
\end{equation}
has to be vanishing for all values of $r$. Eq.~(\ref{overlap1}) is
vanishing if and only if
\begin{equation}\label{overlapMultiplied}
\frac{\dbrb{v}\dket{\Psi_{\rm out} (r,r)}}{\sqrt{1 - |\lambda(r)|^2}} =
\sum_{n = 0}^{\infty} d_{nn}^{\ast} \lambda (r)^n
\end{equation}
vanishes for all $r$. The sum on the right-hand side of
Eq.~(\ref{overlapMultiplied}) can be seen as a polynomial in $\lambda
(r)$ and should vanish for all possible values of $\lambda (r)$,
i.e. for all $|\lambda (r)| < 1$.  Polynomials of this type on a
finite interval form a vector space with linearly independent basis
vectors $\lambda (r)^k$, with $k \in {\mathbbm N}$. Thus the sum in
Eq.~(\ref{overlapMultiplied}) vanishes $\forall r \in \left\langle 0,
  \infty \right)$ only if $d_{nn} = 0$, $\forall n\in {\mathbbm
  N}$. This is in contradiction with our assumption about the vector
$\ket{v}$ and therefore the largest support an operator $\ED$,
unambiguously indicating the difference of the squeezing parameters,
can have is the orthocomplement of the span of vectors $\ket{n}
\ket{n}$, with $n \in {\mathbbm N}$. This concludes the proof.
\par
In the case (ii) of compared states with completely arbitrary phases of the complex squeezing parameters, the proof can be done in the same way as before, up to defining accordingly the set of pairs of same/different states.
\section{Performances of the setup}\label{s:perform}
In this section we give a thorough analysis of the statistics of our
setup also in the presence of non-unit quantum efficiency at the
detection stage in order to assess its reliability in Section
\ref{s:reliab}.
\subsection{Probability of revealing the difference}
\begin{figure}[tb!]
\includegraphics[width=0.75\columnwidth]{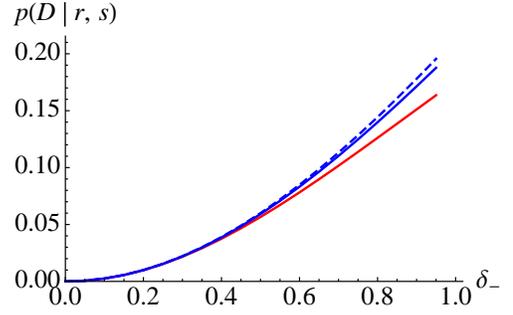}
\caption{(Color online) Plot of the conditional probability of
  revealing the difference of two squeezed vacuum states $\ket{\psi_1}
  = S(r)\ket{0}$ and $\ket{\psi_2} = S(s)\ket{0}$ in the ideal case
  ($\eta=1$) as a function of $\delta_- = |r-s|$. Solid
  lines, from top to bottom, correspond to the optimal squeezed-states
  comparator (blue) and the universal comparator (red line). The
  dashed line is the upper bound on the probability in the case of
  only two possible squeezings. See the section \ref{s:reliab} for
  details. \label{f:OptComp}}
\end{figure}

The conditional probability of revealing the difference of compared
states with $\xi\neq \zeta$ [but $\arg(\xi)=\arg(\zeta)=\varphi$,
though unknown], that is the probability to obtain a $\ED$
outcome, reads:
\begin{equation}\label{p:D}
p(D|\xi,\zeta) = 1- p(0|\xi,\zeta)
\end{equation}
with:
\begin{align}\label{p:inc}
p(0|\xi,\zeta) &= \dbra{\Psi_{\rm out}(\xi,\zeta)}
\EO \dket{\Psi_{\rm out}(\xi,\zeta)} \nonumber\\
&= \left[1-|\lambda(r_{+})|^2\right]
\sum_{n,m=0}^{\infty}[\lambda(r_{+})]^{n}[\lambda^{*}(r_{+})]^{m} \nonumber\\
&\hspace{0.5cm}\times \sum_{k=0}^{\infty}
\big\{[S(r_{-})]_{kn}\big\}^2\, \big\{[S^{\dag}(r_{-})]_{mk}\big\}^2,
\end{align}
where $\dket{\Psi_{\rm out}(\xi,\zeta)}$ is given in
Eq.~(\ref{eq:PsiOutReal}). For $\xi\rightarrow\zeta$ we
correctly obtain $p(0|\xi,\xi)=1$.
By noting that \cite{p:91}:
\begin{align}\label{sq:elem:phase}
  [S(\gamma)]_{hk} \propto \left\{\begin{array}{ll}
  \exp\{i(\frac{h-k}{2})\theta\} & \hbox{for } h,k \hbox{ odd or even},\\
0 & \hbox{otherwise},
\end{array}\right.
\end{align}
where $\gamma = |\gamma| e^{i\theta}$, it is straightforward to see
that Eq.~(\ref{p:inc}) does not depend on the (equal) phase $\varphi$
of $\xi$ and $\zeta$.  Thus, in order to investigate the performances
of the optimal squeezed-states comparator, we may set $\varphi=0$ and
let $\xi=r$ and $\zeta=s$, with $r,s \in \mathbbm{R}$, without loss of
generality. Furthermore, it is possible to show by numerical means
that the probability $p(D|r,s)$ does not depend on the sum
of the squeezing parameter $\delta_+=r+s$, but only on the difference $\delta_- =
|r-s|$. In Fig.~\ref{f:OptComp} we plot the probability $p(D|r,s)$
given in Eq.~(\ref{p:inc}) as a function of $\delta_- = |r-s|$, and we
compare it with the possible use of the universal comparator
\cite{comp1}, which works unambiguously for all pure states leading to
\begin{align}p_{\rm UC}(D|\omega) = \frac12 (1-\omega^2)\,,\end{align}
  where $\omega = |\langle \psi_1 |\psi_2 \rangle| = (\cosh
  \delta_-)^{-1/2}$
  is the overlap between the two squeezed vacuum states.
%
\subsection{Influence of non-ideal detectors}
In a realistic scenario, in which the photon-number resolving
detectors have non-unit quantum efficiency $\eta$, we should modify
the POVM by replacing to the projectors $\ket{n}\!\bra{n}$ in
Eq.~(\ref{POVM:comp}) by the following operators \cite{kk64,lui73}:
\begin{equation}\label{POVM:eta}
\Pi_n(\eta) = \eta^n \sum_{k=n}^{\infty} (1-\eta)^{k-n} \binom{k}{n} \ket{k}\bra{k},
\end{equation}
namely (we assume that the two detectors have the same quantum
efficiency):
\begin{subequations}
\begin{align}
\EO(\eta) &= \sum_{n=0}^{\infty} \Pi_n(\eta)\otimes \Pi_n(\eta),\\
\ED(\eta) &= \one - \EO(\eta).
\end{align}
\end{subequations}
The performance of this kind of detector and its reliability to
resolve up to tens of photons have been recently investigated in Ref.
\cite{all:PRA:10}.Eq.~(\ref{POVM:eta}) shows that the single projector
is turned into a (infinite) sum of projectors. This could be a
relevant issue for protocols that rely on the discrimination between
even and odd number of photons \cite{and:PRA:06}, as we mentioned in
section \ref{s:intro}, since it becomes challenging to detect the
actual parity of the number of incoming photons, $\Pi_n(\eta)$ being a
sum over both even and odd number of photons. For what concerns our
setup, as one may expect, the presence of non unit quantum efficiency
no longer guarantees the unambiguous operation.
\par
In principle, the effect of an imperfect detection could be taken into
account while designing the comparison procedure. However, this would
be mathematically challenging and most probably would not provide an
unambiguous procedure anyway, because of the form of the noise in
realistic detectors. Alternatively, one could try to maximize the
reliability (confidence) of the outcomes (for maximum confidence in
state discrimination see \cite{croke}), nevertheless this would
require to make some particular choice of the prior probabilities
$z_S$ and $z_D$ and of the probability distributions $p_{S}(\Phi)$,
$p_{D}(\Phi)$ (see section \ref{s:opt}). By following this type of
approach, an optimal comparison of coherent states in realistic
conditions can be improved by employing a linear amplifier
\cite{amplif}. On the other hand, as we are going to show in the next
section, the reliability of the difference detection of our proposal
is quite close to unambiguity if the detector efficiency is high
enough.
\par
The conditional probability $p_{\eta}(D|\xi,\zeta)$ for the
  detectors with non-unit quantum efficiency $\eta$ reads:
\begin{equation}\label{p:D:eta}
p_{\eta}(D|\xi,\zeta) = 1- p_{\eta}(0|\xi,\zeta),
\end{equation}
with:
\begin{align}\label{p:inc:eta}
p_{\eta}(0|\xi,\zeta) &=
\dbra{\Psi_{\rm out}(\xi,\zeta)} \EO(\eta) \dket{\Psi_{\rm out}(\xi,\zeta)} \nonumber\\
&\hspace{-1cm}= \left[1-|\lambda(r_{+})|^2\right] \sum_{n,l,m=0}^{\infty}
\eta^{2n}\,[\lambda(r_{+})]^{l}\,[\lambda^{*}(r_{+})]^{m}\nonumber\\
&\hspace{-0.5cm}\times\sum_{h,k=n}^{\infty} (1-\eta)^{h+k-2n}\binom{h}{n}\binom{k}{n}\nonumber\\
&\hspace{-0.5cm}\times
[S(r_{-})]_{kl}\, [S(r_{-})]_{hl}\,[S^{\dag}(r_{-})]_{mk}\, [S^{\dag}(r_{-})]_{mh},
\end{align}
that, in the case of $\xi=\zeta$, reduces to:
\begin{align}
p_{\eta}(0|\xi,\xi) &=
\dbra{\Psi_{\rm out}(\xi,\xi)} \EO(\eta) \dket{\Psi_{\rm out}(\xi,\xi)} \nonumber\\
&\hspace{-1cm}= [1-|\lambda(\xi)|^2] \sum_{n=0}^{\infty}\eta^{2n} |\lambda(\xi)|^{2n}\nonumber\\
&\times {}_2F_{1}[1+n,1+n,1,(1-\eta)^2|\lambda(\xi)|^2],\label{p:xi:xi}
\end{align}
where ${}_2F_{1}$ are hypergeometric functions and $[S (r_{-})]_{lm}$
are the matrix elements of the squeezing operator as in
Eq.~(\ref{final:DPC}).  Because of Eq.~(\ref{sq:elem:phase}), the
probabilities (\ref{p:D:eta}) and (\ref{p:inc:eta}) are still
independent of the unknown value of $\varphi$, thus, from
now on, we set $\varphi=0$ and put $\xi=r$ and $\zeta=s$, with $r,s
\in \mathbbm{R}$, without loss of generality.
\begin{figure}[tb!]
\includegraphics[width=0.75\columnwidth]{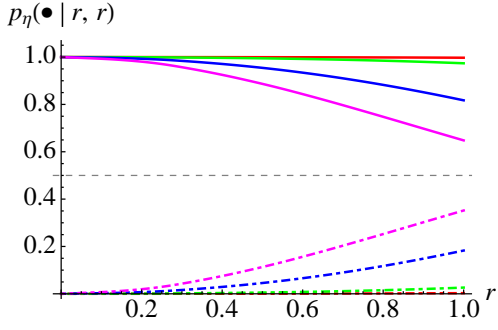}
\caption{(Color online) Plot of $p_{\eta}(0|r,r)$ (solid) and
  $p_{\eta}(D|r,r)$ (dot-dashed lines) as functions of $r$ for
  different values of the efficiency $\eta$; from top to bottom
  (solid) and from bottom to top (dot-dashed lines): $\eta = 0.999$
  (red), $0.99$ (green), $0.90$ (blue), $0.50$
  (magenta). \label{f:p10eta}}
\end{figure}
In Fig.~\ref{f:p10eta} we plot $p_{\eta}(0|r,r)$ and $p_{\eta}(D|r,r)$
for different values of $\eta$. If $r\ll 1$, then Eq.~(\ref{p:xi:xi})
can be expanded up to the second order in $r$, obtaining:
\begin{equation}
p_{\eta}(0|\xi,\xi) \approx 1 - 2 \eta (1-\eta) r^2.
\end{equation}
\subsection{Reliability of the setup}\label{s:reliab}
In order to assess the reliability of our setup, we address the
scenario in which only two squeezing parameters for each of the
squeezed vacua are possible. In such case one knows that the two
squeezing parameters are either $\{(r, r), (s, s)\}$ or $\{(r, s), (s,
r)\}$ with the same prior probability. Our squeezed-states comparator
may not be optimal in this case. However, as one can see in
Fig.~\ref{f:OptComp}, the performance of our setup is nearly as good
as if it was optimized also for this restricted scenario. In
particular, the dashed line in Fig.~\ref{f:OptComp} refers to the
optimal measurement, unambiguously detecting the difference in the
case of only two possible squeezing parameters, in formula \cite{comp3}:
\begin{align}p_{\rm max}(D|\omega)=\frac{
1-\omega^2}{1+\omega^2}\,.\end{align}
\par
\begin{figure}[tb!]
\includegraphics[width=0.75\columnwidth]{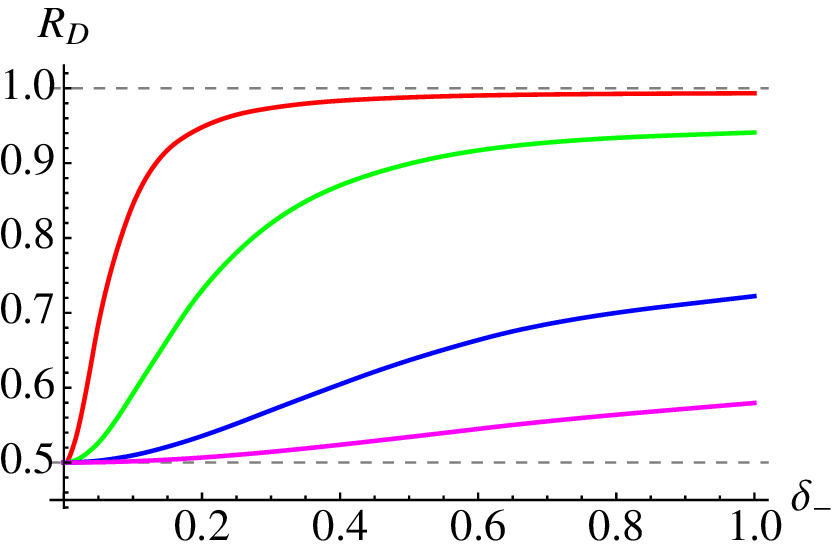}
\includegraphics[width=0.75\columnwidth]{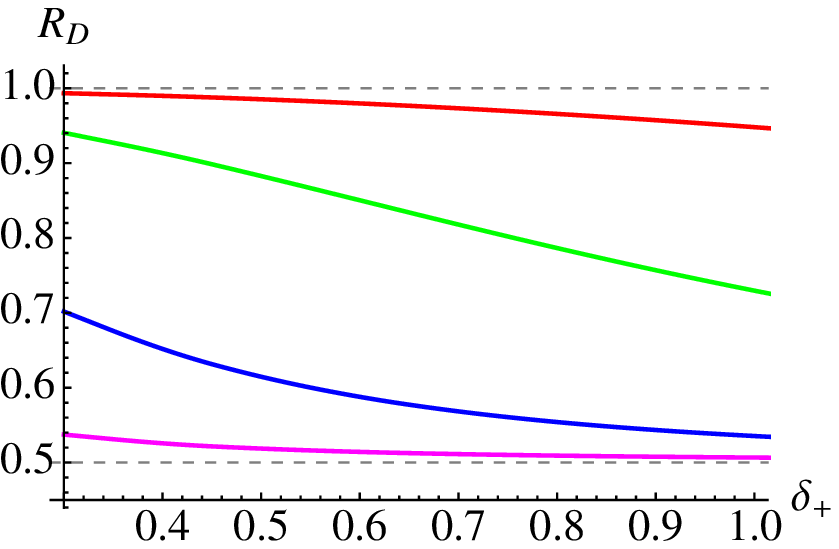}
\caption{(Color online) Top: Reliability $\RD(\eta;r,s)$ (reliability) as a
function of $\delta_-$ for fixed $\delta_+=1.0$ and different values
of the efficiency.  Bottom: Reliability $\RD(\eta;r,s)$ as a
function of $\delta_+$ for difference $\delta_-=0.2$ and different
values of the efficiency. In both plots,
from top to bottom: $\eta = 0.999$ (red), $0.99$
(green), $0.90$ (blue), $0.50$ (magenta). \label{f:RD}}
\end{figure}
We define the reliability $\RD$ of the scheme in revealing the
difference of the squeezing parameters $r$ and $s$ as the conditional
probability of the two squeezed vacuum states being different
if the outcome $\ED$ is found, i.e.,
(we assume equal prior probabilities):
\begin{align}
\RD (\eta;r,s) 
  &= \frac{\, p_\eta \left( D | r,s \right)
  + \,p_\eta \left( D | s,r \right)}
{\sum_{u,v=r,s}\,p_\eta \left( D | u,v \right)}.
\end{align}
In the ideal case, i.e., $\eta=1$, we have $p_\eta\left( D | r,r
\right)=0$ and, thus, $\RD(\eta;r,s)=1$, which is guaranteed by the
construction of the setup. On the other hand, if $\eta < 1$, then
$p_\eta\left( D | r,r \right)\ne 0$ and, consequently, the conclusion
based on the outcome $D$ is not unambiguous anymore. The actual value
of $\RD$ can be numerically calculated starting from
Eq.s~(\ref{p:D:eta}) and (\ref{p:inc:eta}). The reliability
$\RD(\eta;r,s)$ is plotted in the upper panel of Fig.~\ref{f:RD} as a
function of $\delta_- = |r-s|$. Note that differently from the case
$\eta=1$, for $\eta <1$ the probability $p_\eta \left( D | r,s
\right)$ depends not only on the difference $\delta_-=|r-s|$ but also
on the sum $\delta_+=r+s$. The dependence on $\delta_+$ is shown in
the the lower panel of Fig.~\ref{f:RD}, where we plot $\RD(\eta;r,s)$
as a function of $\delta_+$ for fixed difference $\delta_-=0.2$.
\section{Concluding remarks}\label{s:concl}
In this paper we have addressed the comparison of two squeezed vacuum states of which we have a single copy available. We have suggested an optical setup based on a beam splitter, a phase shifter and two photodetectors which is feasible with the current technology. Even though we analyzed the scenario with an equal, though unknown, phase of the compared states, our setup is able to operate unambiguously with ideal detectors irrespective of the squeezing phases, and without the knowledge of the relative phases of the squeezed states. We have proved the optimality of our scheme for arbitrary phases and ideal detectors and we analyzed its performance and reliability also in the presence of non-unit quantum efficiency at the detection stage in the case of equal phases. As one may expect, the detection efficiency strongly affects the reliability; nevertheless we have shown that, for small energies and not too low quantum efficiency, the setup is still robust.
\par
Our scheme may be employed not only for the comparison of two squeezed
vacua, but for a more general scenario in which the input states
$\ket{\xi}$ and $\ket{\zeta}$ are known to be transformed by two {\em
  fixed known} local unitaries $U$ and $V$, respectively (namely, $U
\ket{\xi} \otimes V \ket{ \zeta}$) or by any {\em fixed known} global
unitary transformation $W$ ($W \ket{\xi} \otimes \ket{ \zeta}$): now
it is enough to apply the inverse of the transformation before
processing the state with our setup.
\section*{Acknowledgments}
Fruitful discussions with M.~Ziman are acknowledged.  This work has
been supported by the project INQUEST APVV SK-IT-0007-08 within the
``Executive programme of scientific and technological co-operation
between Italy and Slovakia'', by the European Union projects Q-ESSENCE
248095, HIP 221889, and partially supported by the CNR-CNISM
agreement.
\appendix
\section{Proof of the equivalence of the two schemes}\label{app:CMevol}
In this Appendix we show the equivalence between the schemes in
Fig.~\ref{f:schemes}~(a) and \ref{f:schemes}~(b).  Since the squeezed
states are Gaussian states and all operations involved in the schemes
(phase shift and beam splitter mixing) preserve the Gaussian
character, we use the phase-space description of the system evolution
\cite{FOP:05}. For the sake of simplicity we focus on the case of real
squeezing parameters, i.e., $\xi=r$ and $\zeta=s$, with $r,s \in
{\mathbbm R}$.  The symplectic transformation associated with the
squeezing operator $S (r)$ is:
\begin{equation}
  \label{eq:SqueezingOperatorSymplMatrix} \bmSt (r) = \left( \begin{array}{cc}
    e^r & 0\\
    0 & e^{- r}
  \end{array} \right),
\end{equation}
while the symplectic transformation associated with the balanced beam
splitter operator $\Ubs$ is:
\begin{equation}
\Sbs = \frac{1}{\sqrt{2}} \left(
\begin{array}{cc}
\Id_2 & -\Id_2  \\
\Id_2 & \Id_2
\end{array}
\right),
\end{equation}
where $\Id_2$ is a $2 \times 2$ identity matrix. The covariance matrix
of the outgoing Gaussian state in the scheme Fig.~\ref{f:schemes}~(a)
[for the sake of simplicity we used $U(\pi/2) S(s) = S(-s)$ and we do
not write explicitly the symplectic transformation of the phase
shift]:
\begin{equation}
  \dket{\psiOUT(r,s)} = \Ubs S (r) \otimes S (-s) \ket{0},
\end{equation}
is, thus, given by:
\begin{equation}
  \label{sigma:ev} \bmsigmaOUT = \Sbs \hspace{0.25em} \bmLS (r, -s)
  \hspace{0.25em} \bmsigma_0 \hspace{0.25em} \bmLS (r, -s)^T \hspace{0.25em}
  \Sbs^T,
\end{equation}
where $\bmsigma_0=\frac{1}{2}\Id_4$,
\begin{equation}
  \bmLS (r, -s) = \left( \begin{array}{cc}
    \bmSt (r) & \tmmathbf{0}\\
    \tmmathbf{0} & \bmSt (-s)
  \end{array} \right),
\end{equation}
represents the two local squeezing operations $LS(r, -s) = S(r)
\otimes S(-s)$. The explicit form of (\ref{sigma:ev}) reads:
\begin{equation}
  \label{sigma:out}
  \bmsigmaOUT = \frac{1}{2} \left(\begin{array}{c}
    \begin{array}{cccc}
      f (r, -s) & 0 & g (r, -s) & 0\\
      0 & f (- r, s) & 0 & g (- r, s)\\
      g (r, -s) & 0 & f (r, -s) & 0\\
      0 & g (- r, s) & 0 & f (- r, s)
    \end{array}
  \end{array}\right),
\end{equation}
where:
\begin{eqnarray}
  f (x, y) = \frac{e^{2 x} + e^{2 y}}{2} & \tmop{and} & g (x, y) = \frac{e^{2
  x} - e^{2 y}}{2} .
\end{eqnarray}
Note that by setting $s = r$ one obtains the covariance matrix of the
TWB in Eq.~(\ref{TWB}).
\par
It is now straightforward to verify that the same result of the
evolution as in Fig.~\ref{f:schemes}~(a), corresponding to the
covariance matrix in Eq.~(\ref{sigma:out}), may be obtained
considering the setup displayed in Fig.~\ref{f:schemes}~(b). Here two
input states with same squeezing parameter amplitude $r_{+} = (r + s)
/ 2$ are mixed after a phase shift at the BS and the outgoing modes
undergo two local squeezing operations with amplitude $r_{-} = (r - s)
/ 2$; in formula:
\begin{equation}\label{sigma:ev:2}
\bmsigma' = \bmLS (r_{-}, r_{-}) \hspace{0.25em} \bmS_2 (r_{+})
  \hspace{0.25em} \bmsigma_0 \hspace{0.25em} \bmS_2 (r_{+})^T \hspace{0.25em}
  \bmLS (r_{-}, r_{-})^T,
\end{equation}
where $\bmS_2 (r_{+}) = \Sbs \hspace{0.25em} \bmLS (r_{+}, -r_{+})$ is
the symplectic transformation associated with $S_2 (r)$ defined in
Eq.~(\ref{TWB}). By performing the calculation one finds
$\bmsigma'=\bmsigmaOUT$, and, since Gaussian states are completely
characterized by their covariance matrix (and first moments), one can
conclude that the final states are the same.

\vspace{0.3cm}

\section{No unambiguous detection of sameness of two states}
\label{app:nosame}
In this Appendix we show that the no-error condition given in
Eq.~(\ref{noerrorcond1:b}), together with continuity of the involved
mappings, imply that we cannot unambiguously detect the sameness of
two states. Let us consider a state $\ket{\Phi}=S(r
e^{i\varphi})\ket{0}\otimes S(s e^{i\varphi})\ket{0} \in
\mathcal{S}^{\varphi}_{D}$ with $r\neq s$. The no-error condition
(\ref{noerrorcond1:b}) demand that:
\begin{equation}
\Tr(\PS\ket{\Phi}\bra{\Phi})=0, \quad \forall r\neq s.
\label{app:noerror1} \end{equation}
Let us now take the limit
$s\rightarrow r$. Thanks to continuity of the trace and the chosen
parameterization of the set of states, we conclude that $\forall r$:
\begin{equation}
\bra{0}S^\dagger(r e^{i\varphi})\otimes S^\dagger(r e^{i\varphi}) \PS
S(r e^{i\varphi})\otimes S(r e^{i\varphi})\ket{0}=0.
\end{equation}
It follows that Eq.~(\ref{app:noerror1}) has to hold for arbitrary $r$
and $s$. Since $\PS$ is a positive operator, it should be zero on the
relevant part of the Hilbert space spanned by
$\mathcal{S}^{\varphi}\otimes\mathcal{S}^{\varphi}$, i.e., all the
possible pairs of the compared states. Hence, without loss of
generality, we can choose $\PS=0$ on the whole Hilbert space.

\end{document}